\documentstyle[12pt]{article}
\input epsf
\def\bea{\begin{eqnarray}}
\def\eea{\end{eqnarray}}
\begin{document}
\title{Fading out  of $J/\psi$ color transparency
in high energy  heavy ion peripheral collisions}
\author{
L. Frankfurt\\
\it School of Physics and Astronomy, Raymond and Beverly Sackler\\
\it Faculty of Exact Science, Tel Aviv University, Ramat Aviv 69978,\\
\it Tel Aviv , Israel\\
M. Strikman\\
\it Pennsylvania State University, University Park, Pennsylvania 16802\\
M. Zhalov\\
\it Petersburg Nuclear Physics Institute, Gatchina 188350, Russia}
\date{}
\maketitle
\centerline {\bf ABSTRACT}
We provide predictions for the $J/\psi$
coherent 
production in the peripheral heavy ion collisions at LHC and RHIC using the
leading twist model of nuclear shadowing based on the QCD
factorization theorem for diffraction and the HERA hard diffraction
data. We demonstrate that for LHC kinematics this model leads to
a bump-shape
distribution in rapidity 
which is
suppressed overall as compared to the
expectations of the color transparency regime by a factor $\sim 6$.
This is a significantly larger suppression than
that  expected within
the impact parameter eikonal  model.
Thus we show that the interaction of spatially small wave package
for which the total cross section of interaction with 
 nucleons is small is still strongly shadowed  by nuclear medium
in high energy processes.

\section{Introduction}
Interaction of small size
color singlet objects with hadrons is one  of the most actively
studied issues in high-energy QCD. In perturbative QCD  (similar to QED)
the total cross section of the interaction of such systems with
hadrons is proportional to the area
occupied by color within projectile hadron
\cite{Low} leading to the expectation of
a  color transparency phenomenon for various hard processes with nuclei.
In the case of incoherent processes cross sections are expected to be
proportional to the number of nucleons in the nuclei
while in the case of coherent processes the amplitude
is proportional to number of nucleons times the nuclear form factor.
Possibility to approximate projectile heavy quarkonium as colorless
dipole of heavy quarks can be formally derived from QCD within the limit when
mass of heavy quark $m_{Q}\rightarrow \infty$ but $x_{Bj}=4m_{Q}^2/\nu$
is fixed and not extremely small \cite{FKS}. 
In this kinematics the size of heavy
quarkonium is sufficiently small to justify applicability of PQCD.

Recently the color transparency (CT) phenomenon was observed
at FNAL by E791 experiment\cite{E791} which studied  the coherent
process of dissociation of a 500 GeV  pion
into two jets off the nuclei. The measurement has confirmed a number of
predictions of \cite{FMS} including the A-dependence, and the transverse
and longitudinal momentum distributions of the jets.
Previously the color transparency type behaviour of the cross section
was observed also in the coherent $J/\psi $ photoproduction
at $\left<E_{\gamma}\right>=120 GeV$ \cite{Sokoloff}.

A natural question is whether the color transparency will hold
for arbitrary high energies? Two phenomena are expected to work
against CT at high energies leading to onset of a new regime
which we refer to as {\it the  color opacity regime}. One is the
leading twist gluon shadowing. Indeed the QCD factorization for hard
exclusive coherent processes with nuclei like
$\gamma^*_L+A \to ``Vector~ meson''+A$ implies that
the cross sections are proportional to the square of the
gluon parton density $G_A(x,Q^2)$ at small x which is screened
in nuclei as compared to  the nucleon: $G_A(x,Q^2)/A G_N(x,Q^2) < 1$.
This obviously should lead to a
gradual
disappearance of color transparency \cite{FMS,BFGMS}. Another
mechanism for violation of CT at high energies is the increase of the
small dipole-nucleon cross section with energy $\propto G_{N}(x,Q^2)$.
For sufficiently large energies this cross section becomes
comparable to the meson-nucleon cross sections and hence one may expect
a significant  suppression  of the hard exclusive diffractive processes
like DIS diffractive production of vector meson and photoproduction of
heavy quarkonium states as compared to the CT scenario. However
it seems that this phenomenon is beyond the kinematics achievable
for the photoproduction of $J/\psi$ mesons at RHIC
($x\approx 2\cdot 10^{-2},Q^2\approx 10 GeV^2$)
and probably even at LHC.

It was suggested in \cite{FKS,FS99} to look for color opacity phenomenon
using $J/\psi $ (photo) electroproduction. This however requires energies
much larger than those available at the fixed target facilities and
would require use of electron-nucleus colliders. At  the same time
estimates of the counting rates performed within the framework of the
FELIX study \cite{FELIX} have demonstrated that the effective photon
luminosities  generated in peripheral heavy ion collisions
at LHC would lead to significant  rates of coherent photoproduction
of vector mesons including $\Upsilon$  in reaction
\begin{equation}
 A + A \to A + A + V.
\label{react}
\end{equation}
As a result it would be possible to study at LHC 
photoproduction of vector mesons in Pb-Pb and Ca-Ca collisions
at energies much higher than the range $W_{\gamma p}\le 17.3 GeV$ covered 
at the fixed target 
experiment at FNAL \cite{Sokoloff}. 
Note that even current  experiments at RHIC
($W_{\gamma p}\le 25 GeV$)
should also  exceed this limit.

Currently the theory of photoinduced processes in AA collisions is
well developed, for the recent review
see \cite{baur}. Hence we can combine it with our previous
studies of the coherent photo(electro)production
of vector mesons to make predictions  for production of $J/\psi$ in
the process (\ref{react}). Typical transverse momenta which are
exchanged between two nuclei in  the
peripheral collisions which leave nuclei intact are much smaller than the
typical transverse momenta in the coherent photoproduction
of vector mesons. As a result for the cross section integrated over
the momentum of the nucleus which emits the quasireal photons we can use
the standard Weizsacker-Williams approximation.

Hence  the  cross section of the vector meson production
integrated over the transverse momenta of the nucleus which emitted a
photon  can be written in the convoluted form:
\begin{equation}
{d \sigma(AA\to J/\psi AA)\over dk}=2\int d^2b T_{AA}({\vec b})\frac
{n(k,{\vec b})} {k} \sigma_{\gamma A\rightarrow J/\psi A}(k).
\label{base}
\end{equation}
Here $k=\gamma k_3$ is the photon momentum in
the colliding frame($k_3$ - momentum of photon in the rest
system of emitting nucleus and $\gamma$ - Lorentz factor).

For the quantity  $n(k,{\vec b})$ presenting the flux of photons
with momentum $k$ in the collider frame we used the simplest approximated
form\cite{baur}:
\begin{equation}
n(k,{\vec b}))=\frac {Z^2\alpha} {{\pi}^2} \frac{1} {b^2}X^2
\bigl [K^2_1(X)+\frac {1} {\gamma} K^2_0(X)\bigr ],
\end{equation}
where $K_0(X)$ and $K_1(X)$ are modified Bessel functions with
argument $X=\frac {bk} {\gamma}$ and ${\vec b}$ is the impact 
parameter distance
between centres of colliding nuclei. 
The factor $T_{AA}({\vec b})$ accounts for inelastic interactions of
the nuclei at impact parameters  $b \le 2R_A$. It can be approximately
calculated as
\begin{equation}
T_{AA}({\vec b})=\int d^2b_1T_A({\vec b_1})T_A({\vec b}-{\vec b_1}),
\end{equation}
where $T_A({\vec b})=\int \limits^{\infty}_{-\infty}dz$
$\rho_A(z,{\vec b})$  is the usual profile function of the nucleus.
In our calculations we use the nuclear matter density
$\rho_A(z,{\vec b})$ obtained from the mean field
Hartree-Fock-Skyrme(HFS) model, which describes  many global properties
of nuclei as well as the intermediate and high energy 
elastic proton-nucleus scattering 
and nucleus electromagnetic form factors.
This indicates that the HFS model provides
a good description of both proton and neutron distribution in nuclei
and takes into account a small difference between the matter
distribution and the charge distribution.

The main subject of our interest in this paper is estimating  the
cross section of the process $\gamma A\rightarrow J/\psi A$
\begin{equation}
\sigma_{\gamma A\rightarrow J/\psi A}(k)=  \int dt
{d \sigma(\gamma A\to VA)(\tilde{s},t)\over dt},
\label{phocs}
\end{equation}
where
\begin{equation}
\tilde{s}=4E_N k=4\gamma k m_N
\label{stilde}
\end{equation}
is invariant energy for $\gamma - N$
scattering ($E_N=\gamma m_N$ is the energy per nucleon
in the c.m.  of the nucleus-nucleus  collisions),
$t=-|\vec{p}^V_t|^2$ is square of the vector meson transverse momentum.

\section{Coherent photoproduction of $J/\psi$ off nuclei}

\begin{figure}
\begin{center}
        \leavevmode
        \epsfxsize=.50\hsize
       \epsfbox{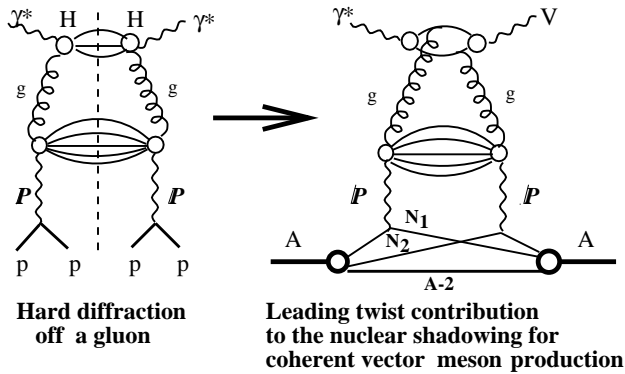}
    \end{center}
\vspace*{1cm}
\caption{Correspondence between diagrams for hard gluon induced
diffraction off nucleon and shadowing for the vector meson production.}
\label{diag0}
\end{figure}

Let us discuss the photoproduction amplitude
$\gamma + A\to J/\psi +A$ in more details.
We are interested here in the $W_{\gamma p}$ range which 
can be probed at RHIC and LHC.
In this situation interaction of $c\bar c$ which in the final state forms
$J/\psi$ is still rather far from the black body limit in which
cross section can be calculated  in the model independent way \cite{BBL}.
Several mechanism of coherent interaction with several nucleons were
suggested    for this process. We focus here on the
 {\bf the leading twist mechanism of shadowing.}
There exist  qualitative difference between
the mechanism of interaction of a small dipole with several  nucleons
and the case of a similar interaction of an ordinary hadron.
Let us for example consider interaction with two nucleons.
The leading twist contribution is described by the diagrams
where two gluons are attached to the
dipole. To ensure that nucleus remains intact in such a process we need to
attach colorless lines to both nucleons.
These diagrams are closely related to the diagrams corresponding to
the gluon diffractive parton densities which are measured
at HERA (see Fig.
~\ref{diag0})   
and hence to the
similar diagrams for the gluon nuclear shadowing \cite{FS99}.

\begin{figure}
\begin{center}
        \leavevmode
        \epsfxsize=.80\hsize
       \epsfbox{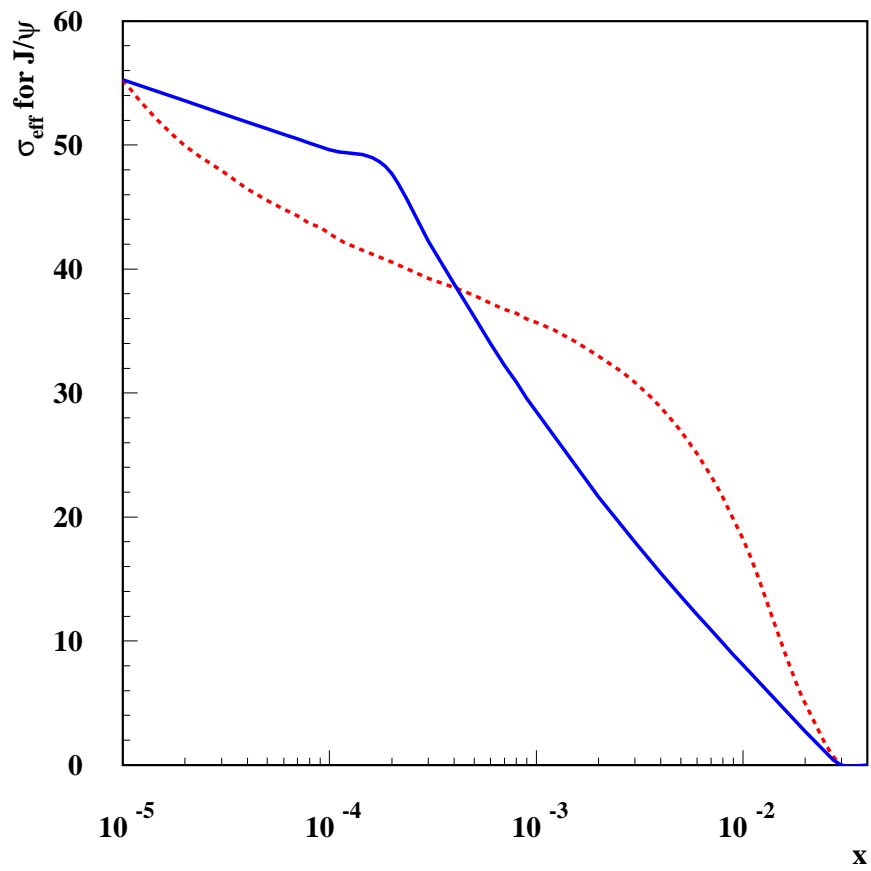}
    \end{center}
\vspace*{1cm}
\caption{
The quantity $\sigma_{eff}^g$ for $Q^2=4 GeV^2$ as a function of the
Bjorken x
for H1 (solid line) and Alvero et al \cite{Alvero} (dashed line)
parameterizations of the gluon diffractive density.}
\label{sigef}
\end{figure}

\begin{figure}
\begin{center}
        \leavevmode
        \epsfxsize=.50\hsize
       \epsfbox{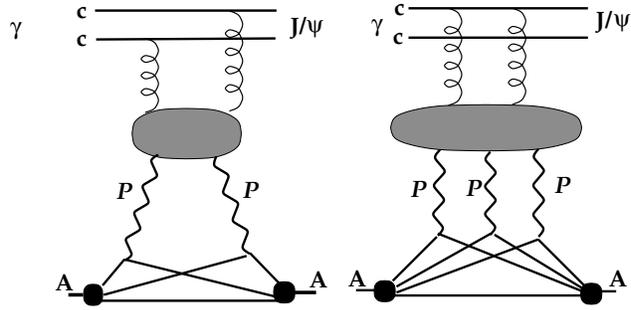}
    \end{center}
\vspace*{1cm}
\caption{Leading twist diagrams for the   production of
$J/\psi$ off two and three nucleons.}
\label{multi}
\end{figure}

As a result
it was possible to express the  quark and gluon nuclear shadowing
for the interaction with two nucleons in a model independent
way through the corresponding diffractive parton densities using
the Gribov theory of inelastic shadowing\cite{Gribovinel} and
the QCD factorization theorem for the hard diffraction
\cite{Collins}.  An important discovery of HERA is that
hard diffraction  is indeed dominated by the leading twist contribution and
gluons play a very important role in the diffraction(this is loosely
referred to as gluon dominance of the Pomeron). Analysis of the
DESY diffractive data indicates that in the gluon induced processes
probability of the diffraction
is much larger than in the quark induced processes
\cite{FS99}. The recent H1 data on diffractive dijet production
\cite{H1} provide an additional confirmation of this observation.
Large probability of diffraction in the gluon induced hard processes
could  be understood in the s-channel language as formation of
color octet dipoles  of rather
large size which can elastically scatter with a rather large cross
section. The strength of this interaction can be quantified using
optical theorem and introducing
\begin{equation}
\sigma_{eff}^g=
16 \pi {d \sigma_{diff}(x, Q^2)/dt(t=0) \over \sigma_{tot}(x,Q^2)}
\end{equation}
for the hard process of scattering of a virtual photon off the gluon
field of the nucleon. An important feature of this mechanism of
coherent interaction is that it is practically absent for
$x\ge 0.02\div 0.03$ and may rather quickly become important with
decrease of $x$. The gluon virtuality scale
which is relevant for the $J/\psi$ photoproduction
is 3-4 GeV$^2$ with a significant fraction of the amplitude
due  to smaller virtualities \cite{FKS,FMS2000}.  Hence we will take
the gluon shadowing in the leading twist at $Q^2=4 GeV^2$. Taking a smaller
value of $Q^2$ would result in even  larger shadowing effect.
We present numerical values of
$\sigma_{eff}^g(Q^2=4 GeV^2)$ for two current models of
the diffractive gluon densities(Fig.~\ref{sigef}) which practically cover
the range of parameterizations
available in the literature. The H1 parameterization leads to
a more graduate onset of the contribution of the
double interactions because in this model diffraction into masses
with $M^2/Q^2\le 1 $ is smaller.
The dijet data of H1 prefer this scenario though it seems that
further measurements will be necessary to
clarify the issue. So we keep both models for the further analysis.
For a more detailed discussion of the current models of diffraction
and of the resulting values of $\sigma_{eff}^g$ see \cite{FGMS}.
The effective cross section
$\sigma_{eff}$ can be used to estimate relative importance of the
interactions with  $N\ge 3$ \cite{FS99}, which corresponds to account of
diagrams of Fig. ~\ref{multi} in the quasieikonal approximation.
As a result the t-dependence of the photoproduction turns out about
the same as for the case of Glauber scattering of a projectile with cross
section of interaction with a nucleon equal to $\sigma_{eff}$.
The ratio of the photoproduction cross sections off nucleus and nucleon is
expressed in the leading twist through the ratio of the skewed gluon
parton densities. In the case of $J/\psi $ production they
are pretty close to the gluon density calculated at
$x_{eff}=(x_1+x_2)/2$ where $x_1, x_2$ are light cone fractions carried
by exchanged gluons and $x=x_1-x_2$.
In the DIS limit, or for $m_Q\to \infty$, one finds \cite{FMS2000}
$x_2 \ll x$ and hence $x_{eff}\approx x/2$. However in the case of $J/\psi$
photoproduction Fermi motion effects lead to $x_2/x \sim 0.3-0.5$ and
hence to $x_{eff}\approx x$.

\begin{figure}
\begin{center}
        \leavevmode
        \epsfxsize=.50\hsize
       \epsfbox{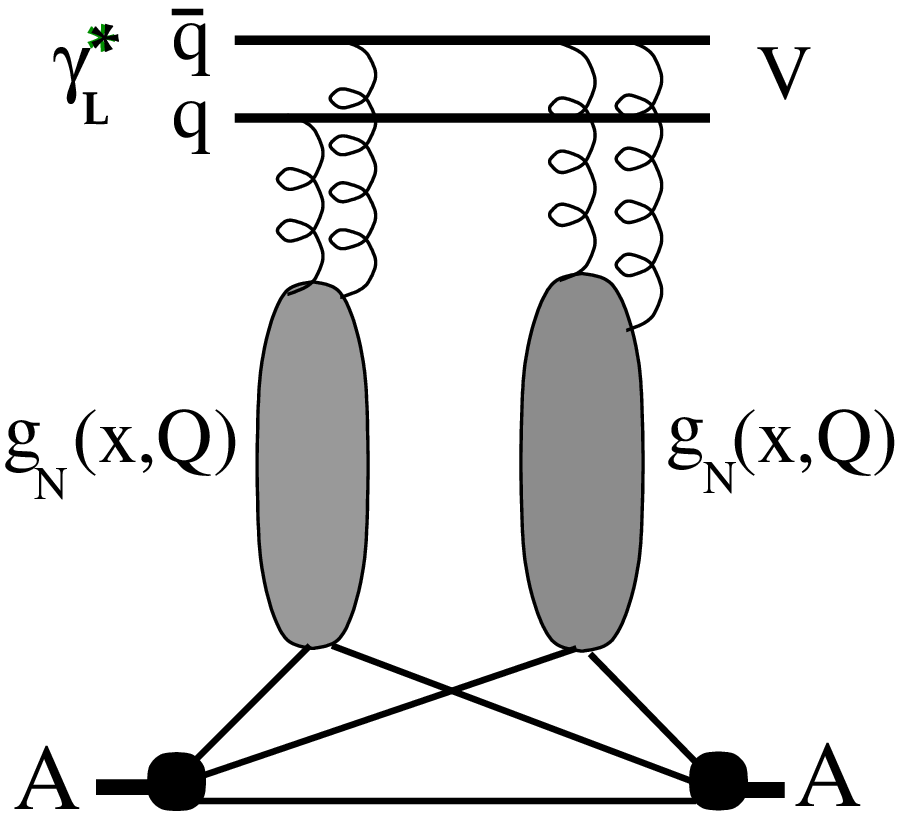}
    \end{center}
\vspace*{1cm}
\caption{Typical diagrams for the   higher twist
eikonal interactions  of a small dipole with two nucleons.}
\label{diag1}
\end{figure}

In principle
the multiple eikonal type rescatterings (at fixed
transverse separations) due to multiple
gluon exchanges - see Fig. ~\ref{diag1} 
({\bf the impact parameter eikonal rescattering model}) could also result in
 suppression of the vector meson production.
Though validity of this approximation is hard to justify 
in QCD the model calculations suggest that this effect is
not small numerically \cite{FKS}. However,  it is still significantly smaller
than the leading twist shadowing (at least for $x \le 0.001$)
which we find in our calculations. Note in passing
that if one would consider the gluon shadowing using phenomenological
models \cite{FLS,eskola} where shadowing for gluons was assumed to be
equal to that for quarks at low normalization scale
one would find comparable suppressions due to the leading twist gluon
shadowing and the eikonal rescatterings. This would make
extraction of the gluon shadowing from  measurements of
photoproduction of $J/\psi$ highly problematic.

\section{Numerical results and discussion.}

Having build the quasieikonal model for the amplitudes of the
interaction of
$c\bar c$ pair with several nucleons we can now calculate the
amplitude of scattering off nuclei.
We have demonstrated in \cite{FS99} that the amplitude in this
approximation has the same structure of the rescattering terms as the
Glauber model with $\sigma_{tot}$ substituted by $\sigma_{eff}$.
Hence we can use the optical limit of  the Glauber model\cite{yenn}
to calculate the cross section of $J/\psi$ photoproduction

\begin{eqnarray}
 {d \sigma_{\gamma A\to VA}(\tilde{s},t)\over dt}=
 {d \sigma_{\gamma N\to VN}(\tilde{s},t=0)\over dt}
{%\vert 
\left|
\int d^2bdz e^{i{\vec q_t}\cdot {\vec b}}\rho ({\vec b},z)
e^{iq_lz}\cdot e^{-\frac {1} {2} \sigma _{eff}( {M^2_V}/\tilde{s})
\int
\limits ^{\infty}_{z}
\rho({\vec b},z')dz'} 
%\vert
\right|
}^2.
\label{dsig}
\end{eqnarray}

Here the exponential factor with  $q_l=m_V^2/2k$ accounts
for finite longitudinal distances in the transition $\gamma\to V$
(finite longitudinal momentum transfer). The forward elementary cross
section for photoproduction of $J/\psi$ meson on
nucleon was taken using the fit to experimental
data presented in \cite{landshoff} (this is preferable to using 
the theoretical
calculations which for photoproduction of $J/\psi$ 
have theoretical uncertainty of the order of two).

\begin{figure}
\begin{center}
        \leavevmode
        \epsfxsize=.80\hsize
       \epsfbox{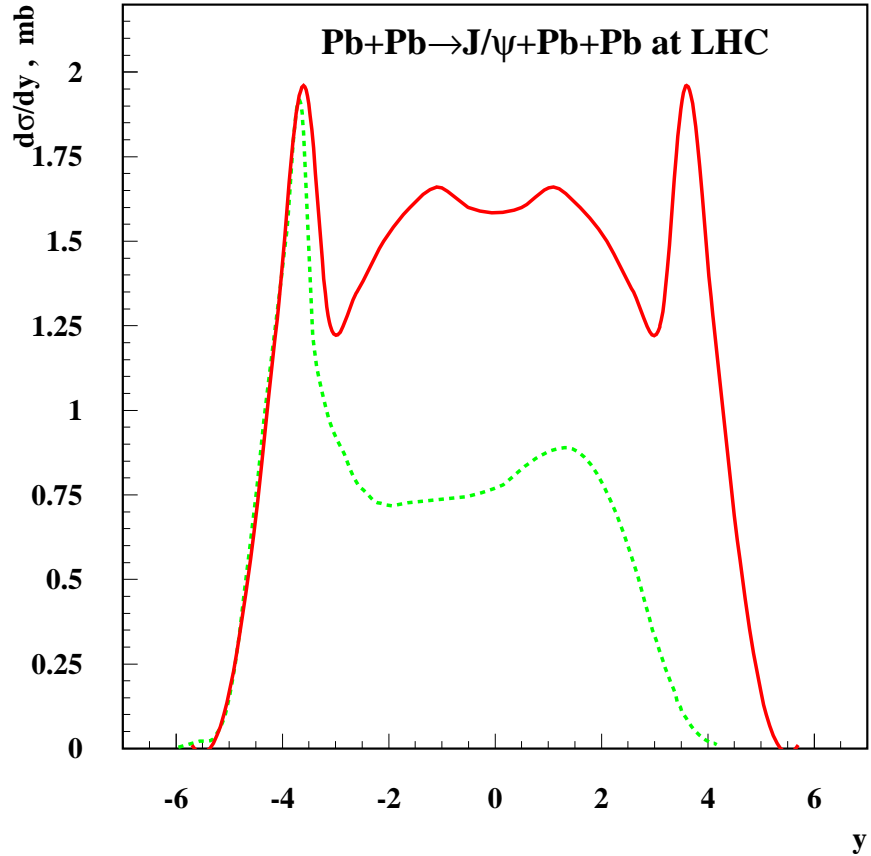}
    \end{center}
\vspace*{1cm}
\caption{
The rapidity distribution for the $J/\psi$ production in lead-lead
peripheral collisions at LHC. Solid line - production by two-side beams, 
dashed - production by one-side beam only. Calculations were
performed with $\sigma^{J/\psi N}_{eff}$ based on Alvero et.al.
 parameterization of the gluon diffractive density.}
\label{lhcpbpb1}
\end{figure}

\begin{figure}
\begin{center}
        \leavevmode
        \epsfxsize=.80\hsize
       \epsfbox{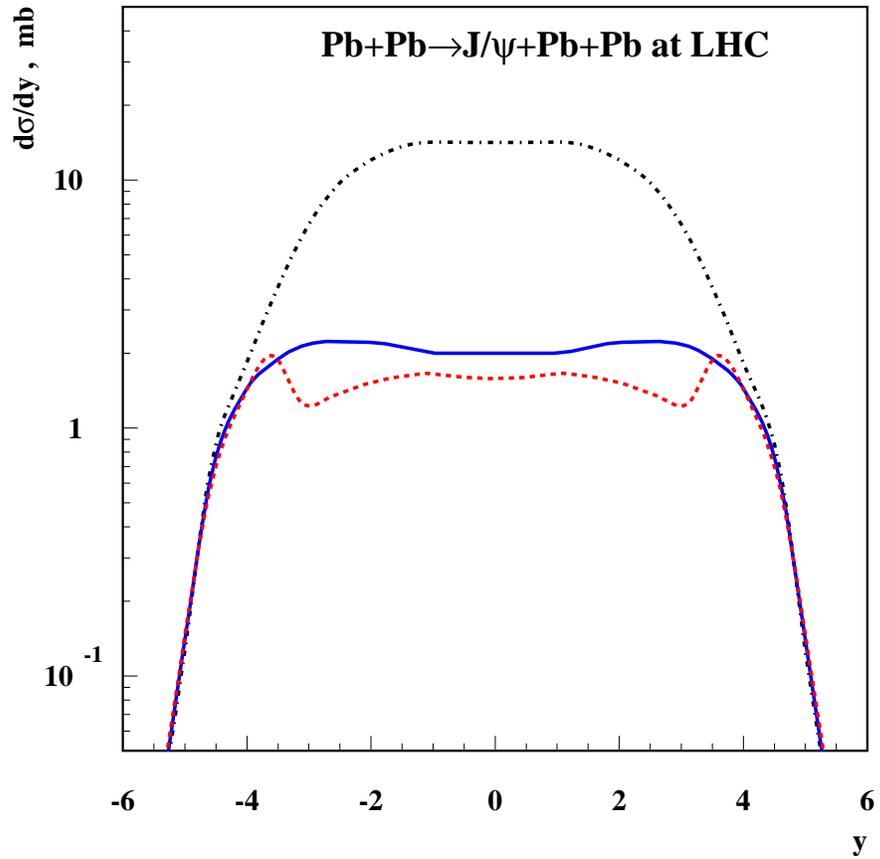}
    \end{center}
\vspace*{1cm}
\caption{
The rapidity distributions for the LHC lead-lead peripheral collision
$J/\psi$ coherent production calculated with  $\sigma_{eff}^{J/\psi N}$
based on H1(solid line) and  Alvero et al (dashed line) parameterizations
of gluon density and in the impulse approximation (dot-dashed line)}
\label{lhcpbpb2}
\end{figure}

\begin{figure}
\begin{center}
        \leavevmode
        \epsfxsize=.80\hsize
       \epsfbox{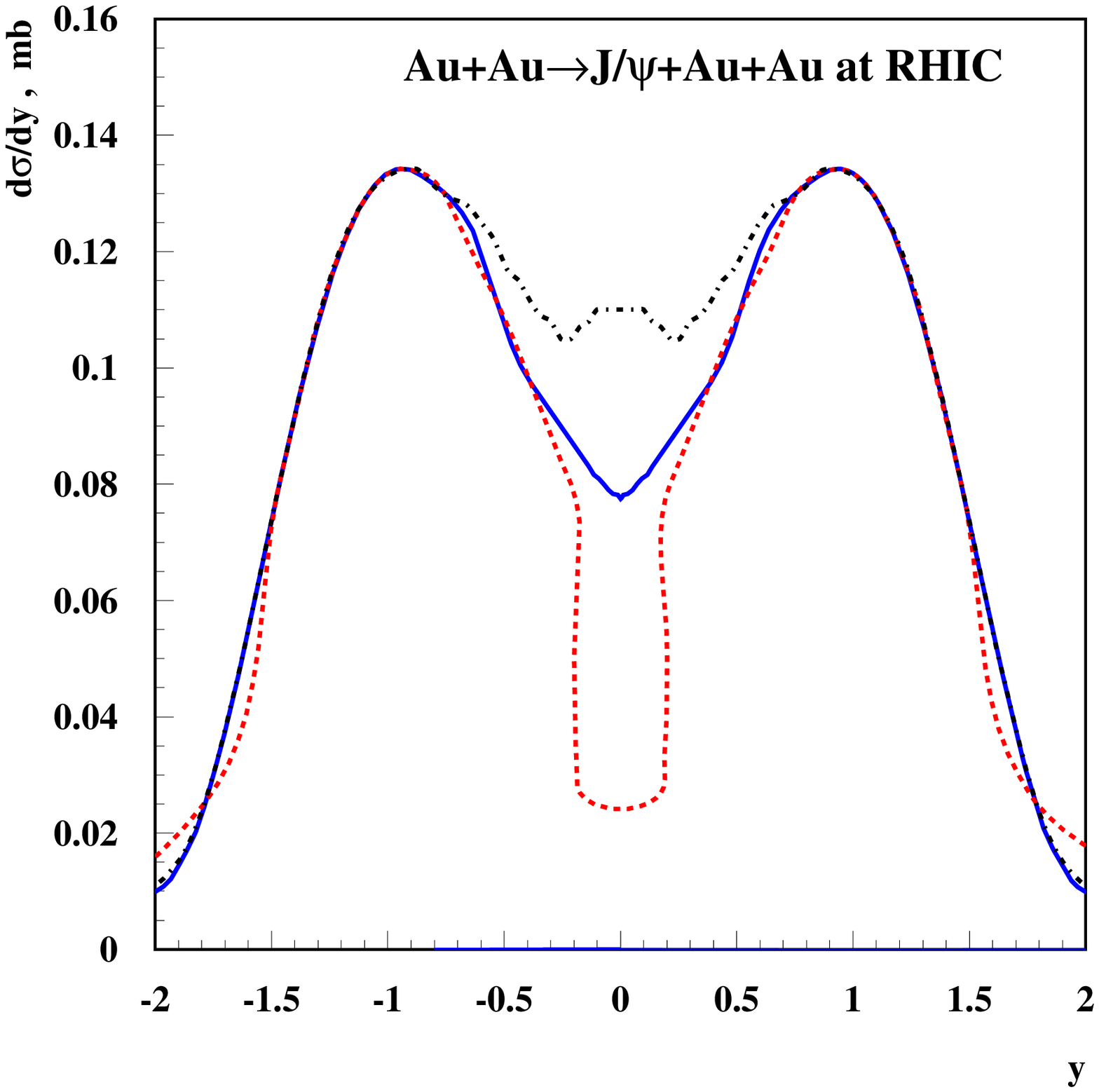}
    \end{center}
\vspace*{1cm}
\caption{
The same as in Fig.~\ref{lhcpbpb2} but for gold-gold collisions at RHIC.}
\label{rhicauau}
\end{figure}

We focus here on the distributions over rapidity:
\begin{equation}
y={1\over 2}\ln{E_V-p_3^V \over E_V +p_3^V}=\ln {2k\over m_V}.
\end{equation}
In  Figs.~\ref{lhcpbpb1}, ~\ref{lhcpbpb2}, and  Fig.~\ref{rhicauau} we
present the differential cross sections
both including effects of gluon shadowing and
without gluon shadowing (impulse approximation)
for lead-lead peripheral collisions at LHC and gold-gold
collisions at RHIC.

One can see that
on the top of the overall suppression of the cross section
the gluon  nuclear shadowing leads to
a significant modification of the
shape of  the rapidity distribution as compared to the
impulse approximation.
Bumps near  the edges  of rapidity distribution are due to
a  sharp increase of the effective cross section found in the
calculation based on the Alvero et. al. model of the gluon diffractive
density in the region of
 Bjorken ${\it x}$ close to ${\it x\approx 10^{-2}}$. A bump in the
center region of $\it y$
 arises due to the drop of the photon flux as can be  seen from
 Fig.~\ref{lhcpbpb1}. These effects
are weakened for the H1 parameterization which leads to a more
gradual increase of $\sigma_{eff}$
and they disappear in the impulse approximation.
It is also of interest that in the LHC kinematics we are sensitive
to the cross section of
photoproduction at $W_{\gamma p}$ up to a factor of three
 larger than $W_{\gamma p}$
corresponding to production at $y=0$. Hence the measurements will
actually probe the $J/\psi$ photoproduction at the energies beyond
those reachable at HERA in electron-nucleus mode.

In the RHIC kinematics we find even more
nice picture
in the case of gold-gold collision.
The decrease of cross section as a function of rapidity due to the
shadowing is combined
with drop of the photon flux in the same region of rapidities. This
results in a narrow dip at
$y=0$ which is very sensitive to pattern of onset of the gluon
shadowing.  The test of this prediction will be feasible at RHIC
since the rates of $J/\psi $ production are pretty high \cite{Klein}.

At LHC the total cross section calculated with accounting for
the gluon shadowing effects is
$\sigma(PbPb\to J/\psi+PbPb) \approx 14$ mb  and the value
$\sigma(PbPb\to J/\psi+PbPb) \approx 85$ mb was obtained in the impulse
approximation. Hence, we have  found a strong suppression
of the $J/\psi$ yield for this case as it was predicted on the base of
rough estimates in\cite{FS99}.
In kinematic of RHIC the effect of suppression due to the
gluon shadowing is rather small
and $\sigma(AuAu\to J/\psi+AuAu) \approx 0.320$ mb while in
the impulse approximation
$\sigma(AuAu\to J/\psi+AuAu) \approx 0.360$ mb.

Let us compare our results with two other calculations of the reaction (1).
The first rather detailed calculation of the coherent process
$AA\rightarrow A+V+A$ has been reported in Ref.\cite{Klein}. To evaluate
nuclear shadowing effects in the total $J/\psi A$ cross section the vector
dominance model,  classical mechanics formulae (accounting for the elastic
rescatterings of vector meson only) have been used in \cite{Klein}.
On the contrary our calculation uses eikonal approximation where
inelastic shadowing effects dominate. Really $\sigma^g_{eff}$ derived
from the diffractive gluon densities includes both the elastic and
inelastic shadowing. It is also assumed in Ref.\cite{Klein} that
the $t$-dependence of the cross section is $\propto |F_{A}(t)|^2$
(where $F_{A}(t)$ is the nuclear form factor) while account for the
rescattering effects (eq.~\ref{dsig}) leads to somewhat steeper t-dependence.
For the RHIC energies whenever comparison is possible the results of
Ref.\cite{Klein} are pretty close to ours. This is because nuclear
shadowing effects are a small correction for the photoproduction of
$J/\psi$ in the kinematics of RHIC. For the LHC kinematics we obtained
cross section for  the coherent $J/\psi$ production significantly below
the value of \cite{Klein} (a factor of two for lead -lead collisions).
The difference is because approach used in \cite{Klein} significantly
underestimates the strengths of multiple interaction of $c\bar c$ pair
with the nucleus. Also rapidity distributions for lead-lead collisions for
which we find an interesting shape were not considered in
Ref.\cite{Klein}.

After this study was nearly completed a report has appeared \cite{BG}
where it has been suggested to use the coherent $J/\psi$ production in
the peripheral ion-ion collisions to measure shadowing of gluon densities
in nuclei. The analysis in \cite{BG} is based on the factorization
theorem of \cite{FMS,BFGMS}, the $t$-dependence of the coherent
$\gamma A\to J/\psi A$ cross section has been approximated as $F^2(t)$
and three sets of the gluon distributions has been used. Calculations with
the GRV gluon distribution is effectively equivalent to the impulse
approximation. Two others model accounts for the nuclear shadowing. The
distribution of \cite{eskola} assumes the same shadowing for gluons as
for quarks which is in variance with diffractive data from HERA. The
second model \cite{AyalaFilho:2001cq} attempts to account for nonlinear
QCD evolution in the gluon density and in this case application of the
factorization approximation is hardly justified.
It would be reasonable to expect that at least for $J/\psi$
production in the kinematics at RHIC where the shadowing effects don't
influence essentially  the total cross sections
our numerical results should be qualitatively similar.
However we found gross differences - our absolute cross sections
are considerably larger as compared to that in \cite{BG} both
for the kinematics at RHIC and at LHC.

To conclude,
we have demonstrated that heavy ion collisions at
RHIC are sensitive to the onset of gluon nuclear shadowing while
the measurements at LHC will allow to establish disappearance
of transparency of nuclear matter for spatially small
$c\bar c$  dipole
at high energies .

We thank R.Engel for  useful discussions and GIF, CRDF and DOE for support.

\end{document}